\documentclass[english]{article}
\usepackage[T1]{fontenc}
\usepackage[latin1]{inputenc}
\usepackage{verbatim}
\usepackage{amsmath}
\usepackage{amssymb}

\makeatletter

\let\SF@@footnote\footnote
\def\footnote{\ifx\protect\@typeset@protect
    \expandafter\SF@@footnote
  \else
    \expandafter\SF@gobble@opt
  \fi
}
\expandafter\def\csname SF@gobble@opt \endcsname{\@ifnextchar[
  \SF@gobble@twobracket
  \@gobble
}
\edef\SF@gobble@opt{\noexpand\protect
  \expandafter\noexpand\csname SF@gobble@opt \endcsname}
\def\SF@gobble@twobracket[#1]#2{}
\providecommand{\tabularnewline}{\\}

\newcommand{\lyxrightaddress}[1]{
\par {\raggedleft \begin{tabular}{l}\ignorespaces
#1
\end{tabular}
\vspace{1.4em}
\par}
}

\usepackage{geometry}
\usepackage{mcite}
\makeatother

\usepackage{babel}
\makeatother
\begin{document}

\title{Pseudoclassical description of scalar particle in non-Abelian background
and path-integral representations}

\author{R. Fresneda%
\thanks{fresneda@fma.if.usp.br%
} and D. Gitman%
\thanks{gitman@dfn.if.usp.br%
}}

\maketitle
\begin{abstract}
Path-integral representations for a scalar particle propagator in
non-Abelian external backgrounds are derived. To this aim, we generalize
the procedure proposed by Gitman and Schvartsman 1993 of path-integral
construction to any representation of $SU\left(N\right)$ given in
terms of antisymmetric generators. And for arbitrary representations
of $SU\left(N\right)$, we present an alternative construction by
means of fermionic coherent states. From the path-integral representations
we derive pseudoclassical actions for a scalar particle placed in
non-Abelian backgrounds. These actions are classically analyzed and
then quantized to prove their consistency. 
\end{abstract}

\lyxrightaddress{Insituto de Física, Universidade de São Paulo, São Paulo, Brasil.}

\section{Introduction}

QFT with external backgrounds is a good approach for describing many
physical situations and effects. If the external background is strong
enough it has to be taken into account non-perturbatively. The corresponding
methods for QED are well developed and were fruitfully applied for
a number of calculations, see e.g. \cite{qed-lebedev,*grib1994,*Fradkin:1991zq,*Greiner:1985ce}
and citations therein. The external background concept in non-Abelian
QFT is less developed and meets some difficulties (there is no gauge
invariant way of introducing a non-Abelian external field). However,
the undeniable existence of physical situations where there is a sufficiently
strong quantized non-Abelian field often serves as a physical justification
for treating this field as an external classical field, in spite of
the above mentioned problem. Interesting physically meaningful results
obtained in this conceptual framework serve as an additional justification
for it. We can point out calculations of one-loop effective actions
in constant non-Abelian external fields \cite{Ambjorn:1983ne,*Ambjorn:1982nd,*Ambjorn:1982bp,Vainshtein:1984xd,Reuter:1996zm}
that were used for constructing the true QCD vacuum, see \cite{Reuter:1996zm,Savvidy:1977as,*Nielsen:1978rm,*Ragiadakos:1980ah,Yildiz:1979vv,*Claudson:1980yz,*Adler:1981sn,*Dittrich:1983ej,*Flory:1983dx,*Cho:2000ck,Blau:1988iz}.
One also ought to mention the description of phase-transitions in
cosmological QCD \cite{Bhattacharyya:1999dm}, non-pertubative parton
production from vacuum by a classical $SU\left(3\right)$ \cite{Nayak:2005yv}
and $SU\left(2\right)$ \cite{Kharzeev:2006zm} chromoelectric field,
boundary conditions and topological effects of the vacuum in the presence
of a non-homogenous external magnetic field in the form of a flux
tube \cite{Serebryanyi:1986uk,Gornicki:1990kq}, and so on.

The key objects in nonperturbative (with respect to the background)
QFT with a non-Abelian background are scalar and spinning particle
propagators in the corresponding non-Abelian external field. Exact
solutions for such objects allow one to obtain by an integration one-loop
results for various physical quantities. Moreover, path-integral representations
for the propagators may be useful in obtaining exact solutions, which
could then be used in calculations. Manifold path-integral representations
for scalar and spinning particle propagators were constructed and
calculated for various Abelian backgrounds in \cite{barbashov65,*Batalin:1970it,*Henneaux:1982ma,Borisov:1982cp,Fainberg:1987jr,*Polyakov:1987ez,*Marshakov:1988zp,*Fainberg:1988zg,*Fainberg:90,*Ambjorn:1989ba,*Grundberg:1989pg,*Korchemsky:1989sz,*Grundberg:1990ai,Fradkin:1991ci,Gitman:1992an,Gitman:1993yb,Korchemsky:1992ru,*Aliev:1994vxa,*Holten:1995ds,*Karanikas:1995ua}.
It turned out that such representations are also useful for deriving
the so-called pseudoclassical actions for spinning particles, see
\cite{Fradkin:1991ci,Gitman:1992an,Gitman:1996wk,*Geyer:1999xp}.
Some path-integral representations for propagators in non-Abelian
backgrounds and problems related to the pseudoclassical description
of isospin were studied in \cite{Balachandran:1976ya,Barducci:1976xq,Borisov:1982cp}.
We recall that a classical theory for a Yang-Mills particle was first
constructed from the classical limit of the Yang-Mills field equations
by Wong \cite{Wong:1970fu}. Afterwards, Chen and Dresden \cite{Dresden:1975uv}
showed that the Yang-Mills field equations imply the equations of
motion for a test particle with isotopic spin in a way similar as
the Einstein equations imply the equations of a massive test particle.
Casalbuoni et.al. \cite{Barducci:1976xq} obtained a gauge-invariant
Lagrangian description of scalar and spinning particles with isotopic
spin, where Grassmann variables describe the internal degrees of freedom
at the classical level, so that quantization gives finite-dimensional
representations of the gauge group. Balachandran et.al. \cite{Balachandran:1976ya}
applied Dirac quantization to a pseudoclassical Lagrangian formulation
of scalar and spinning particles interacting with a non-Abelian gauge
field, and additionally developed the method we use here to obtain
the irreducible representations of isospin. In \cite{Gitman:1993yb},
the isospinor structure of the propagator of a scalar relativistic
particle in the fundamental representation of $SU\left(2\right)$
is derived from a path-integral representation using methods developed
for the case of the spinning particle.

In the present article we return once again to these problems for
the case of a scalar particle with isospin placed in various non-Abelian
external backgrounds. We point out that a quantized scalar field in
a non-Abelian background has been put forward as a tentative explanation
of QCD confinement by means of a massive scalar particle (dilaton)
\cite{Dick:1997ta,*Dick:1998hv,*Chabab:2000du}, and also appears
in the form of fundamental scalars coupled to gauge curvature terms
in string theory \cite{Green:1987sp}.

We construct path-integral representations for the scalar particle
propagator from two approaches: one is a generalization of the procedure
proposed in \cite{Gitman:1993yb} to any representation of $SU\left(N\right)$
given in terms of antisymmetric generators, while the other is a constructed
using fermionic coherent states valid for arbitrary representations
of $SU\left(N\right)$. The latter approach is a modification of the
path-integral representation of the Dirac propagator by means of fermionic
coherent states presented in \cite{Borisov:1982cp}. In both cases
we derive the pseudoclassical actions for a scalar particle in non-Abelian
backgrounds, and quantize them to prove their consistency. In the
Appendix, we put some technical details and proofs. The developed
techniques can be easily generalized to the case of a spinning particle
in non-Abelian and gravitational backgrounds. Such a generalization
is the subject of our next publication.

\section{Propagator representations}

The causal propagator for a relativistic scalar particle interacting
with a $su\left(N\right)$ valued external field $\mathcal{A}_{\mu}$
in Minkowski spacetime (in natural units $\hbar=c=1$) is described
by the equation \begin{equation}
\left(\mathcal{\mathcal{P}}^{2}-m^{2}\right)_{\,\,\beta}^{\alpha}D_{\,\,\gamma}^{\beta}\left(x,y\right)=-\delta_{\gamma}^{\alpha}\delta^{4}\left(x-y\right)\,,\,\,\mathcal{\mathcal{P}}_{\mu}=i\partial_{\mu}-q\mathcal{A}_{\mu}\,,\label{flat-propagator-def}\end{equation}
 where $\mathcal{A}_{\mu}=A_{\mu}^{a}t_{a\beta}^{\alpha}$ is a linear
combination of the traceless hermitian matrices $t_{a\beta}^{\alpha}$,
$a=1,..,N^{2}-1$ which are the generators of the Lie algebra $su\left(N\right)$
in an $n\times n$ irreducible matrix representation whose indices
are labeled by greek letters from the beginning of the alphabet, $\alpha$,$\beta$,$\gamma$,etc.,
$\alpha=1,...,N$. Since $SU\left(N\right)$ is a compact group, there
is a basis where its structure constants are totally antisymmetric
and purely imaginary,\begin{equation}
\left[t_{a},t_{b}\right]=f_{ab}^{c}t_{c},\,\, f_{ab}^{c}\equiv f_{\left[abc\right]}\,,\label{eq:lie-algebra}\end{equation}
 and the generators can be normalized as $\mathrm{tr}\left(t_{a}t_{b}\right)=1/2\delta_{ab}$.

In the following we will consider two different realizations of the
Lie algebra (\ref{eq:lie-algebra}) of $su\left(N\right)$. The first
realization will be in terms of creation and annihilation operators
defined on a suitable Fock space, and the second realization will
be in terms of the generators of a suitable Clifford algebra.

I. Let us consider the first realization. Consider an abstract Hilbert
space $\mathcal{H}$ which is the direct product of the usual representation
space for the Heisenberg algebra, whose basis vectors are denoted
as $\left|x\right\rangle $, \begin{align}
 & \hat{x}^{\mu}\left|x\right\rangle =x^{\mu}\left|x\right\rangle \,,\,\,\left\langle x\right.\left|y\right\rangle =\delta^{4}\left(x-y\right)\,,\,\,\int d^{4}x\left|x\right\rangle \left\langle x\right|=I\,,\nonumber \\
 & \left[\hat{x}^{\mu},\hat{p}_{\nu}\right]=i\delta_{\nu}^{\mu}\,,\,\,\left\langle x\right|\hat{p}_{\mu}\left|y\right\rangle =-i\partial_{\mu}\delta^{4}\left(x-y\right)\,,\label{eq:x-basis}\end{align}
 and an abstract Hilbert space $V$ which we do not specify for the
time being, but whose orthonormal basis vectors are $\left|\alpha\right\rangle $,
$\alpha=1,...,n$,\begin{equation}
\left\langle \alpha\right.\left|\beta\right\rangle =\delta_{\alpha\beta}\,,\,\,\sum_{\alpha=1}^{n}\left|\alpha\right\rangle \left\langle \alpha\right|=I\,.\label{eq:fock-basis}\end{equation}
 Thus, the abstract Hilbert space $\mathcal{H}=H\otimes V$ has the
orthonormal basis $\left|x,\alpha\right\rangle =\left|x\right\rangle \otimes\left|\alpha\right\rangle $,
$\left\langle x,\alpha\right.\left|y,\beta\right\rangle =\delta^{4}\left(x-y\right)\delta_{\alpha\beta}$.

Next, we interpret the matrix operators appearing in (\ref{flat-propagator-def}),
as matrix elements of operators in $\mathcal{H}$. With this in mind,
the propagator $D_{\,\,\beta}^{\alpha}\left(x,y\right)$ is the matrix
element of an abstract operator $\hat{D}$, \begin{equation}
D\left(x,y\right)_{\,\,\beta}^{\alpha}=\left\langle x,\alpha\right|\hat{D}\left|y,\beta\right\rangle \,,\label{propagator-matrix-element}\end{equation}
 and the generators $t_{a\beta}^{\alpha}$ are matrix elements of
the operators $\hat{t}_{a}$,\[
\left\langle \alpha\right|\hat{t}_{a}\left|\beta\right\rangle =t_{a\beta}^{\alpha}\,.\]
 We note that if the matrix elements of the operators $\hat{t}_{a}$
are generators of a representation the algebra $su\left(N\right)$,
then so are the operators themselves:\begin{equation}
\left[t_{a},t_{b}\right]_{\,\,\beta}^{\alpha}=f_{ab}^{c}t_{c\beta}^{\alpha}\Leftrightarrow\left[\hat{t}_{a},\hat{t}_{b}\right]=f_{ab}^{c}\hat{t}_{c}\,.\label{eq:induced-representation}\end{equation}
 Using the operators just defined, one can write (\ref{flat-propagator-def})
in operator form,\[
\left(\hat{P}^{2}-m^{2}\right)\hat{D}=-I\,,\]
 where\[
\hat{P}_{\mu}=-\hat{p}_{\mu}-q\hat{\mathcal{A}}_{\mu}\,,\,\,\hat{\mathcal{A}}_{\mu}=A_{\mu}^{a}\left(\hat{x}\right)\hat{t}_{a}\,,\,\,\left\langle x,\alpha\right|\hat{P}_{\mu}\left|y,\beta\right\rangle =\left(i\partial_{\mu}\delta_{\alpha\beta}-qA_{\mu}\left(x\right)t_{a\beta}^{\alpha}\right)\delta^{4}\left(x-y\right)\,.\]
 Thus, one can formally write the inverse of the operator $\hat{D}$,
\[
\hat{D}=-\left(\hat{P}^{2}-m^{2}+i\varepsilon\right)^{-1}\,,\]
 by means of the proper time representation\begin{equation}
\hat{D}=i\int_{0}^{\infty}d\lambda e^{-i\hat{H}\left(\lambda\right)}\,,\,\,\hat{H}=-\lambda\left(\hat{P}^{2}-m^{2}+i\varepsilon\right)\,.\label{eq:propertime}\end{equation}

Let us now further specify $\mathcal{H}$ by defining $V$ as the
one-particle sector of the Fock space for the fermionic creation and
annihilation operators $a^{\dagger}$ and $a$,\[
\hat{a}_{\alpha}\left|0\right\rangle =0\,,\,\,\hat{a}_{\alpha}^{\dagger}\left|0\right\rangle =\left|\alpha\right\rangle \,,\]
 which satisfy the algebra\begin{equation}
\left[\hat{a}_{\alpha}^{\dagger},\hat{a}_{\beta}\right]_{+}=\delta_{\alpha\beta}\,,\,\,\left[\hat{a}_{\alpha}^{\dagger},\hat{a}_{\beta}^{\dagger}\right]_{+}=\left[\hat{a}_{\alpha},\hat{a}_{\beta}\right]_{+}=0\,.\label{eq:CA-operators}\end{equation}
 Then it is possible to represent the operators $\hat{t}_{a}$ as
\begin{equation}
\hat{t}_{a}=\hat{a}_{\alpha}^{\dagger}t_{a\beta}^{\alpha}\hat{a}_{\beta}\,,\,\, t_{a\beta}^{\alpha}=\left\langle \alpha\right|\hat{t}_{a}\left|\beta\right\rangle \,.\label{eq:CA-representation}\end{equation}
 Here it is important to observe that $\hat{t}_{a}$ are generators
of a representation of $su\left(N\right)$,\[
\left[\hat{t}_{a},\hat{t}_{b}\right]=f_{ab}^{c}\hat{t}_{c}\,,\]
 since their matrix elements $t_{a\beta}^{\alpha}$ satisfy the $su\left(N\right)$
commutation relations (\ref{eq:induced-representation}). In addition,
tracelessness and hermiticity of $t_{a\beta}^{\alpha}$ imply the
same for the operators $\hat{t}_{a}$, \begin{align*}
 & \mathrm{tr}\hat{t}_{a}\equiv\sum_{\alpha=1}^{M}\left\langle \alpha\right|\hat{t}_{a}\left|\alpha\right\rangle =t_{a\alpha}^{\alpha}=0\\
 & \hat{t}_{a}^{\dagger}=\left(\hat{a}_{\alpha}^{\dagger}t_{a\beta}^{\alpha}\hat{a}_{\beta}\right)^{\dagger}=\hat{a}_{\beta}^{\dagger}\bar{t}_{a\beta}^{\alpha}\hat{a}_{\alpha}=\hat{t}_{a}\,,\end{align*}
 where the $\dagger$-involution of the abstract operator algebra
complex-conjugates the matrix entries of $t_{a\beta}^{\alpha}$ in
the above. Finally, we note that $\hat{t}_{a}$ conserves the number
of particles. Using the representation (\ref{eq:CA-representation})
for the generators $t_{a}$ and following \cite{Ohnuki:1978jv,Borisov:1982cp},
we now introduce coherent states $\left|\chi\right\rangle $ and $\left\langle \bar{\chi}\right|$
defined by the exponential of the fermion operators $\hat{a}$ and
$\hat{a}^{\dagger}$ acting on the vacuum:\[
\left|\chi\right\rangle =D\left(\chi\right)\left|0\right\rangle \,,\,\,\left\langle \bar{\chi}\right|=\left|\chi\right\rangle ^{\dagger}\,,\,\, D\left(\chi\right)=e^{\hat{a}^{\dagger}\chi-\hat{a}\bar{\chi}}\,,\,\,\left[\hat{a}_{\alpha},D\left(\chi\right)\right]_{-}=\chi_{\alpha}D\left(\chi\right)\,,\]
 where $\chi_{\alpha}$ and $\bar{\chi}_{\alpha}=\chi_{\alpha}^{\dagger}$
are Grassmann numbers that commute with the vacuum state. Consequently,
these states satisfy%
{}{}{}{} \begin{align*}
 & \hat{a}^{\alpha}\left|\chi\right\rangle =\chi^{\alpha}\left|\chi\right\rangle \,,\,\,\left\langle \bar{\chi}\right|\hat{a}_{\alpha}^{\dagger}=\left\langle \bar{\chi}\right|\bar{\chi}_{\alpha}\,,\\
 & \left\langle \bar{\chi}\right|\left.\xi\right\rangle =e^{\frac{1}{2}\left(\chi\bar{\chi}+\xi\bar{\xi}-2\xi\bar{\chi}\right)}\,,\,\,\int\prod_{\alpha=1}^{N}d\bar{\chi}_{\alpha}d\chi^{\alpha}\left|\chi\right\rangle \left\langle \bar{\chi}\right|=\hat{1}_{V}\,,\,\,\int d\chi\chi=\int d\bar{\chi}\bar{\chi}=1\,.\end{align*}
 Using the above identity resolutions, it is possible to relate matrix
elements from the one-particle sector fock-space basis $\left|\alpha\right\rangle $
to the coherent basis $\left|\chi\right\rangle $,\begin{equation}
\left\langle \alpha\right|\cdot\left|\beta\right\rangle =\int\prod_{\sigma,\kappa=1}^{N}d\bar{\chi}_{\sigma}^{\prime}d\chi^{\prime\sigma}d\bar{\chi}_{\kappa}d\chi^{\kappa}e^{\frac{1}{2}\left(\chi^{\prime}\bar{\chi}^{\prime}+\chi\bar{\chi}\right)}\chi^{\prime\alpha}\left\langle \bar{\chi}^{\prime}\right|\cdot\left|\chi\right\rangle \bar{\chi}_{\beta}\,,\label{eq:rep-transformation}\end{equation}
 where we have used $\left\langle \bar{\chi}\right|\left.\alpha\right\rangle =\bar{\chi}_{\alpha}\exp\frac{1}{2}\chi\bar{\chi}$.
As a consequence, we are able to recast the original form of the propagator
(\ref{propagator-matrix-element}), as matrix elements of one-particle
Fock states, in terms of matrix elements of the coherent states,\begin{equation}
D\left(x,y\right)_{\,\,\beta}^{\alpha}=\int\prod_{\sigma,\kappa=1}^{N}d\bar{\chi}_{\sigma}^{\prime}d\chi^{\prime\sigma}d\bar{\chi}_{\kappa}d\chi^{\kappa}e^{\frac{1}{2}\left(\chi^{\prime}\bar{\chi}^{\prime}+\chi\bar{\chi}\right)}\chi^{\prime\alpha}\left\langle x,\bar{\chi}^{\prime}\right|\hat{D}\left|y,\chi\right\rangle \bar{\chi}_{\beta}\,.\label{eq:coherent-rep-propagator}\end{equation}
 In the next section, the matrix elements $\left\langle x,\bar{\chi}^{\prime}\right|\hat{D}\left|y,\chi\right\rangle $
will be used to to obtain a path-integral representation for the propagator.

II. Another possible interpretation of the propagator $D\left(x,y\right)$
appearing in (\ref{flat-propagator-def}) can be simply as the matrix
elements\[
D\left(x,y\right)_{\,\, j}^{i}=\left\langle x\right|\hat{D}_{\,\, j}^{i}\left|y\right\rangle \]
 of the basis elements $\left|x\right\rangle $ of the abstract Hilbert
space $H$. The abstract operator $\hat{D}$ acquires indices directly
from the matrices of the generators of $su\left(N\right)$. Notice
we have relabeled the indices of the matrix representation of $su\left(N\right)$.
The new indices $i$ and $j$ denote the matrix entries of a new set
of generators $T_{a}$,\begin{equation}
T_{a}=\frac{1}{4}\Gamma_{\alpha}t_{a\beta}^{\alpha}\Gamma_{\beta}\,,\,\,\left[\Gamma_{\alpha},\Gamma_{\beta}\right]=2\delta_{\alpha\beta}\,.\label{eq:antisym-gamma-rep}\end{equation}
 These generators are very convenient for obtaining path-integral
representations of the propagator using techniques adapted from the
spinning particle case. However, for $T_{a}$ satisfying (\ref{eq:lie-algebra}),
this is a representation only if the matrices $t_{a}$ are antisymmetric,
$t_{a}^{T}=-t_{a}$. This drawback can be circumvented if we take
the $t_{a}$ matrices in the adjoint representation $t_{ab}^{c}=f_{ab}^{c}$.
Besides, there are be situations where it is possible to choose antisymmetric
$t_{a}$ for different irreducible representations. For instance,
in the case of $SU\left(2\right)$, it is always possible to choose
antisymmetric $t_{a}$ for the integer spin $s$ representations.
In this case, $\alpha,\beta=1,...,2s+1$and $i,j=1,...,2^{s}$. In
the general case, in the adjoint representation, $\alpha,\beta=1,...,N^{2}-1$
and thus $i,j=1,...,2^{\left[\left(N^{2}-1\right)/2\right]}$. In
the familiar case of the adjoint representation of $su\left(2\right)$,
one has \[
T_{i}=\frac{i}{4}\varepsilon_{ijk}\Gamma_{k}\Gamma_{j}=-\frac{i}{4}\varepsilon_{ijk}\Gamma_{j}\Gamma_{k}\,,\,\, i,j,k=1,2,3\,,\]
 where the $\Gamma$'s satisfy $\left[\Gamma_{i},\Gamma_{j}\right]_{+}=2\delta_{ij}$
and are order $2$ matrices, so they can be chosen to be the Pauli
matrices, $\Gamma_{i}=\sigma_{i}$ %
\footnote{$\sigma_{1}=\left(\begin{array}{cc}
0 & 1\\
1 & 0\end{array}\right)\,,\,\,\sigma_{2}=\left(\begin{array}{cc}
0 & -i\\
i & 0\end{array}\right)\,,\,\,\sigma_{3}=\left(\begin{array}{cc}
1 & 0\\
0 & -1\end{array}\right)$%
},\begin{equation}
T_{i}=-\frac{i}{4}\varepsilon_{ijk}\sigma_{j}\sigma_{k}=\frac{1}{2}\sigma_{i}\,.\label{eq:su2-adjoint-rep}\end{equation}
 The generators $T_{i}$ are hermitian and traceless, and they satisfy
the $su\left(2\right)$ algebra\[
\left[T_{i},T_{j}\right]=i\varepsilon_{ijk}T_{k}\,.\]
 This case is special, because the choice of the adjoint representation
for $t_{a}$ gives $T_{a}$ in the fundamental representation. Another
special situation occurs with $SU\left(4\right)$, where we can choose
$t_{a}$ to be antisymmetric matrices of order $6$, since $su\left(4\right)\simeq so\left(6\right)$.
Thus one has even or odd spinors of $so\left(6\right)$ with $4$
components, giving by means of a method described in sections 3.1
and 4.1, the fundamental representation of $SU\left(4\right)$.

\section{Path integral in coherent states representation}

\subsection{Path integral}

Our goal in this section is to write a path-integral representation
for\begin{equation}
D_{\chi}\left(x,\bar{\chi}^{\prime};y,\chi\right)\equiv\left\langle x,\bar{\chi}^{\prime}\right|\hat{D}\left.y,\chi\right\rangle =i\int_{0}^{\infty}d\lambda\left\langle x,\bar{\chi}^{\prime}\right|e^{-i\hat{H}\left(\lambda\right)}\left|y,\chi\right\rangle \label{eq:coherent-propagator}\end{equation}
 We insert $N-1$ identity resolutions $I=\int dxd\bar{\chi}d\chi\left|x,\chi\right\rangle \left\langle x,\bar{\chi}\right|$
and $N$ integration over $\lambda$:\begin{align}
D_{\chi}\left(x,\bar{\chi}^{\prime};y,\chi\right) & =\lim_{N\rightarrow\infty}i\int_{0}^{\infty}d\lambda_{0}\int\left(\prod_{k=1}^{N-1}dx_{k}d\bar{\chi}_{k}d\chi_{k}\right)d\lambda_{1}\cdots d\lambda_{N}\notag\\
 & \prod_{k=1}^{N}\left\langle x_{k},\bar{\chi}_{k}\right|e^{-i\hat{H}\left(\lambda_{k}\right)/N}\left|x_{k-1},\chi_{k-1}\right\rangle \delta\left(\lambda_{k}-\lambda_{k-1}\right)\,,\label{eq:propagator-discretization}\end{align}
 where $x_{N}=x$, $\bar{\chi}_{N}=\bar{\chi}^{\prime}$, $x_{0}=y$
and $\chi_{0}=\chi$. In order to evaluate the general matrix element
appearing in (\ref{eq:propagator-discretization}), one must choose
a definite ordering prescription for the operators in $\hat{H}$.
In particular, one must solve the ordering ambiguity of the four-fermion
term in $\hat{P}^{2}$. In \cite{Borisov:1982cp}, an additional identity
resolution is inserted between the $\hat{P}$ operators as a solution
to the ordering problem. We do not know to which ordering prescription
this corresponds, and conventional ordering prescriptions such as
Weyl ordering and normal ordering are not gauge-invariant. In the
sequel we show that Weyl ordering is not gauge-invariant, and compute
the resulting effective action. As shown in the Appendix (\ref{eq:hamiltonian-wo}),
the Hamiltonian operator differs from the Weyl-ordered%
\footnote{Weyl ordering here means total symmetrization in bosonic degrees of
freedom, and total antisymmetrization in fermionic degrees of freedom.%
} expression by the term $\lambda\frac{q^{2}}{4}\mathrm{tr}\left(t_{a}t_{b}\right)\hat{A}_{\mu}^{a}\hat{A}^{\mu b}$.
This action, apart from the gauge-breaking term, is identical to the
one that would be obtained by doubling the time partition.

Applying the midpoint rule (\ref{eq:fermionic-midpoint}) for the
general matrix elements gives \begin{align*}
\left\langle x_{k},\bar{\chi}_{k}\right|\hat{H}\left(\lambda_{k}\right)\left|x_{k-1},\chi_{k-1}\right\rangle  & =\int\frac{dp_{k}}{\left(2\pi\right)^{4}}d\bar{\eta}_{k}d\eta_{k}\left\langle x_{k},\bar{\chi}_{k}\right|\left.p_{k},\eta_{k}\right\rangle \left(H_{W}\left(\lambda_{k}\right)+Q\left(\lambda_{k}\right)\right)\left\langle p_{k},\bar{\eta}_{k}\right|\left.x_{k-1},\chi_{k-1}\right\rangle \,,\\
 & H_{W}\left(\lambda_{k}\right)\equiv H_{W}\left(\lambda_{k},\frac{x_{k}+x_{k-1}}{2},p_{k},\bar{\eta}_{k},\frac{\eta_{k}+\chi_{k-1}}{2}\right)\,,\\
 & Q\left(\lambda_{k}\right)\equiv\lambda_{k}\frac{q^{2}}{4}\mathrm{tr}\left(t_{a}t_{b}\right)A_{\mu}^{a}\left(\frac{x_{k}+x_{k-1}}{2}\right)A^{b\mu}\left(\frac{x_{k}+x_{k-1}}{2}\right)\,,\end{align*}
 where $H_{W}$ is the Weyl-symbol of $\hat{H}_{W}$. Substituting
the delta functions $\delta\left(\lambda_{k}-\lambda_{k-1}\right)$
by their integral representations and using the integral representations
of the fermionic delta (\ref{eq:coherent-delta}) for the $\chi$
and $\bar{\chi}$integrations, we have\begin{align*}
 & D_{\chi}\left(x,\bar{\chi}^{\prime};y,\chi\right)=\lim_{N\rightarrow\infty}i\int_{0}^{\infty}d\lambda_{0}\int\left(\prod_{k=1}^{N-1}dx_{k}\right)\left(\prod_{k=1}^{N}\frac{dp_{k}}{\left(2\pi\right)^{4}}d\lambda_{k}\frac{d\pi_{k}}{\left(2\pi\right)}d\bar{\eta}_{k}d\eta_{k}\right)\exp\frac{1}{2}\left(\chi^{\prime}\bar{\chi}^{\prime}-\eta_{N}\bar{\eta}_{N}+2\bar{\chi}^{\prime}\eta_{N}\right)\\
 & \exp i\sum_{k=1}^{N}\left\{ p_{k}\frac{\left(x_{k}-x_{k-1}\right)}{\Delta t}+\pi_{k}\frac{\left(\lambda_{k}-\lambda_{k-1}\right)}{\Delta t}-\frac{i}{2}\frac{\left(\eta_{k}-\eta_{k-1}\right)}{\Delta t}\bar{\eta}_{k}-\frac{i}{2}\frac{\left(\bar{\eta}_{k}-\bar{\eta}_{k-1}\right)}{\Delta t}\eta_{k-1}-H_{W}\left(\lambda_{k}\right)-Q\left(\lambda_{k}\right)\right\} \Delta t\,,\end{align*}
 where $H_{W}\left(\lambda_{k}\right)=H_{W}\left(\lambda_{k},\frac{x_{k}+x_{k-1}}{2},p_{k},\bar{\eta}_{k},\frac{\eta_{k}+\eta_{k-1}}{2}\right)$,
$\eta_{0}=\chi$. The term $\chi^{\prime}\bar{\chi}^{\prime}-\eta_{N}\bar{\eta}_{N}+2\bar{\chi}^{\prime}\eta_{N}$
comes from $\left\langle \bar{\chi}^{\prime}\right|\left.\eta_{N}\right\rangle $,
and in the limit $N\rightarrow\infty$ will reduce to $2\bar{\chi}^{\prime}\eta\left(1\right)$.
Taking the limit $N\rightarrow\infty$ $\left(\Delta t\rightarrow0\right)$
and renaming $\eta\rightarrow\chi$ and $\bar{\eta}\rightarrow\bar{\chi}$,
one has\begin{align}
 & D_{\chi}\left(x,\bar{\chi}^{\prime};y,\chi\right)=i\int_{0}^{\infty}d\lambda_{0}\int DxDpD\lambda D\pi D\bar{\chi}D\chi\exp iS_{eff}\exp\bar{\chi}\left(1\right)\chi\left(1\right)\,,\nonumber \\
 & S_{eff}=\int_{0}^{1}dt\left(p\dot{x}+\pi\dot{\lambda}+\frac{i}{2}\left(\bar{\chi}\dot{\chi}-\dot{\bar{\chi}}\chi\right)+\lambda\left(\left(p_{\mu}+qA_{\mu}^{a}I_{a}\right)^{2}-m^{2}\right)-\frac{q^{2}}{4}\lambda\mathrm{tr}\left(t_{a}t_{b}\right)A_{\mu}^{a}A^{b\mu}\right)\,,\label{eq:coherent-pathintegral}\end{align}
 where $I_{a}=\bar{\chi}t_{a}\chi$, and the functional integration
is performed over the paths $x^{\mu}\left(t\right)$, $p_{\mu}\left(t\right)$,
$\lambda\left(t\right)$, $\pi\left(t\right)$, $\bar{\chi}\left(t\right)$
and $\chi\left(t\right)$, with boundary values $x^{\mu}\left(0\right)=y^{\mu}$,
$x^{\mu}\left(1\right)=x^{\mu}$, $\lambda\left(0\right)=\lambda_{0}$,
$\bar{\chi}\left(1\right)=\bar{\chi}^{\prime}$ and $\chi\left(0\right)=\chi$.

Since the path integral is translation-invariant, one can integrate
over the momenta $p_{\mu}$ by shifting $p\mapsto p+\tilde{p}$, where
$\tilde{p}=-\dot{x}/2\lambda-qA^{a}I_{a}$ is the solution to the
classical equation $\dot{x}=\partial H_{eff}/\partial p$. One finds
after making the substitution $2\lambda=e$, the Lagrangian form of
the path integral:\begin{align}
D_{\chi}\left(x,\bar{\chi}^{\prime};y,\chi\right) & =i\int_{0}^{\infty}d\lambda_{0}\int DxDeD\pi D\bar{\chi}D\chi M\left[e,x\right]\exp i\left(S_{eff}+S_{G}\right)\exp\left(\bar{\chi}\left(1\right)\chi\left(1\right)\right)\,,\nonumber \\
 & S_{eff}=\int_{0}^{1}dt\left(-\frac{\dot{x}^{2}}{2e}-\frac{e}{2}m^{2}-q\dot{x}^{\mu}A_{\mu}^{a}I_{a}+\frac{i}{2}\left(\bar{\chi}\dot{\chi}-\dot{\bar{\chi}}\chi\right)\right)\,,\label{eq:lagrangian-pathintegral}\end{align}
 with the Lagrangian measure and reparametrization gauge-fixing term
$S_{G}$ \begin{align}
 & M\left[e,x\right]=\int Dp\exp\frac{i}{2}\int_{0}^{1}e\left(p^{2}-\frac{q^{2}}{4}\mathrm{tr}t_{a}t_{b}A_{\mu}^{a}A^{\mu b}\right)dt\label{eq:lagrangian-measure}\\
 & S_{G}=\int_{0}^{1}\pi\dot{e}d\tau\label{eq:SG}\end{align}
 Thus, the path-integral representation for the propagator can be
derived with an unambiguous ordering prescription (Weyl-ordering)
at the cost of defining a gauge non-invariant measure.

\subsection{Pseudoclassical action}

The action functional $S_{eff}$ in (\ref{eq:lagrangian-pathintegral}),\begin{equation}
S_{eff}=\int_{0}^{1}dt\left(-\frac{\dot{x}^{2}}{2e}-\frac{e}{2}m^{2}-q\dot{x}^{\mu}A_{\mu}^{a}I_{a}+\frac{i}{2}\left(\bar{\chi}\dot{\chi}-\dot{\bar{\chi}}\chi\right)\right)\,,\,\, I_{a}=\bar{\chi}t_{a}\chi\,,\label{eq:coherent-action}\end{equation}
 is reparametrization invariant,\begin{equation}
\delta_{\epsilon}S_{eff}=0\,,\,\,\delta_{\epsilon}x=\epsilon\dot{x}\,,\,\,\delta_{\epsilon}e=\frac{d}{dt}\left(\epsilon e\right)\,,\,\,\delta_{\epsilon}\chi=\epsilon\dot{\chi}\,,\,\,\delta_{\epsilon}\bar{\chi}=\epsilon\dot{\bar{\chi}}\,.\label{eq:reparam-inv}\end{equation}
 In the gauge $e=\sqrt{\dot{x}^{2}}/m$ it coincides with the action
given in \cite{Barducci:1976xq,Balachandran:1976ya} describing a
scalar relativistic particle with anticommuting coordinates in a representation
of a symmetry group $G$, whose equations of motion are\begin{equation}
m\frac{d}{dt}\frac{\dot{x}_{\mu}}{\sqrt{\dot{x}^{2}}}=q\dot{x}^{\nu}F_{\mu\nu}^{a}I_{a}\,,\,\, D_{t}\chi^{\alpha}\equiv\frac{d}{dt}\chi^{\alpha}+iq\dot{x}^{\mu}A_{\mu}^{a}t_{a\beta}^{\alpha}\chi^{\beta}=0\,,\label{eq:eom}\end{equation}
 where $F_{\mu\nu}^{a}=\partial_{\mu}A_{\nu}^{a}-\partial_{\nu}A_{\mu}^{a}+iqf_{bc}^{a}A_{\mu}^{b}A_{\nu}^{c}$
is the field-strength and $D_{t}$ is the covariant derivative.

For canonical analysis purposes%
\footnote{Definitions and conventions are those used in \cite{gitman:1990qh}.%
}, however, it is better to start from the reparametrization invariant
action (\ref{eq:coherent-action}). Since this action does not contain
derivatives of the einbein, it is best to consider it as a velocity
(see \cite{Gitman:2002fd}), and not introduce its conjugate momentum.
One thus arrives at the following Hamiltonian,\[
H=-\frac{e}{2}T-\dot{\chi}_{\alpha}\phi_{\alpha}-\dot{\bar{\chi}}_{\alpha}\bar{\phi}_{\alpha}\,,\]
 where the set of constraints $\Phi=\left\{ T,\phi,\bar{\phi}\right\} $,
\[
T=\left(p_{\mu}+qA_{\mu}^{a}I_{a}\right)^{2}-m^{2}\,,\,\,\phi_{\alpha}=\pi_{\alpha}-\frac{i}{2}\bar{\chi}_{\alpha}\,,\,\,\bar{\phi}_{\alpha}=\bar{\pi}_{\alpha}-\frac{i}{2}\chi_{\alpha}\,,\]
 defines a degenerate supermatrix $\left\{ \Phi,\Phi\right\} $. The
constraint algebra is simplified if we consider an equivalent set
of constraints $\left\{ \tilde{T},\phi,\bar{\phi}\right\} $, where
$\tilde{T}$ is obtained from $T$ through the shifts $\chi\rightarrow\chi-i\bar{\phi}$
and $\bar{\chi}\rightarrow\bar{\chi}-i\phi$,\[
\left\{ \tilde{T},\phi_{\alpha}\right\} =\left\{ \tilde{T},\bar{\phi}_{\alpha}\right\} =0\,,\,\,\left\{ \phi_{\alpha},\bar{\phi}_{\beta}\right\} =-i\delta_{\alpha\beta}\,.\]
 The new Hamiltonian with redefined Lagrange multipliers is \[
\tilde{H}=\Lambda\tilde{T}+\Lambda_{\alpha}\phi_{\alpha}+\bar{\Lambda}_{\alpha}\bar{\phi}\,,\]
 giving the following time-evolution for the constraints, \[
\frac{d}{dt}\tilde{T}=0\,,\,\,\frac{d}{dt}\phi_{\alpha}=i\bar{\Lambda}_{\alpha}\,,\,\,\frac{d}{dt}\bar{\phi}_{\alpha}=i\Lambda_{\alpha}\,,\]
 so the condition of conservations of the constraints in time simply
determines $\Lambda$ and $\bar{\Lambda}$. The equations of motion
for the independent variables $\eta=\left(x^{\mu},p_{\mu},\chi_{\alpha},\bar{\chi}_{\alpha}\right)$
are given by\[
\dot{\eta}=\left\{ \eta,\Lambda\tilde{T}\right\} _{D\left(\phi\right)}\,,\,\,\phi_{\alpha}=\bar{\phi}_{\alpha}=\tilde{T}=0\,,\]
 where the Dirac brackets have been constructed with regard to the
second-class constraint set $\left\{ \phi;\bar{\phi}\right\} $.Using
well known properties of the Dirac brackets, the equations of motion
become\[
\dot{\eta}=\left\{ \eta,\Lambda T\right\} _{D\left(\phi\right)}\,,\,\,\phi_{\alpha}=\bar{\phi}_{\alpha}=T=0\,,\]
 And the nonzero brackets between independent variables are\begin{equation}
\left\{ x^{\mu},p_{\nu}\right\} _{D\left(\phi\right)}=\delta_{\nu}^{\mu}\,,\,\,\left\{ \chi_{\alpha},\bar{\chi}_{\beta}\right\} _{D\left(\phi\right)}=-i\delta_{\alpha\beta}\,.\label{eq:canonical-brackets-I}\end{equation}
 Moreover, the $I_{a}$ are covariantly constant generators of $SU\left(N\right)$,
\begin{equation}
\left\{ I_{a},I_{b}\right\} _{D\left(\phi\right)}=-if_{ab}^{c}I_{c}\,,\,\, D_{t}I_{a}\equiv\frac{d}{dt}I_{a}+iq\dot{x}^{\mu}A_{\mu}^{b}f_{ab}^{c}I_{c}=0\,,\label{eq:isospin}\end{equation}
 hence are called isospin.

From (\ref{eq:canonical-brackets-I}), we see that the Grassmann operators
will generate a creation-annihilation operator algebra,\begin{equation}
\chi_{\alpha}\rightarrow a_{\alpha}\,,\,\,\bar{\chi}_{\alpha}\rightarrow a_{\alpha}^{\dagger}\,,\,\,\left[a_{\alpha},a_{\beta}^{\dagger}\right]_{+}=\delta_{\alpha\beta}\,.\label{eq:operator-algebraI}\end{equation}
 The Hilbert space $\mathcal{H}$ can be realized as the direct product
of a representation space for the Heisenberg algebra and the $2^{n}$-dimensional
Fock space of the creation and annihilation operators, \begin{equation}
\left|x;\alpha_{1}\cdots\alpha_{p}\right\rangle =a_{\alpha_{1}}^{\dagger}\cdots a_{\alpha_{p}}^{\dagger}\left|x;0\right\rangle \in\mathcal{H}\,,\,\, p=0,..,n\,.\label{eq:Hilbert-space-I}\end{equation}
 As is well-known, the group $SO\left(2n\right)$ preserves the commutation
relations (\ref{eq:operator-algebraI}), and the $so\left(2n\right)$
generators in the above representation are given by $c_{\alpha\beta}=\left[a_{\alpha},a_{\beta}^{\dagger}\right]/2$,
$a_{\alpha}a_{\beta}$ and $a_{\alpha}^{\dagger}a_{\beta}^{\dagger}$.
The $c_{\alpha\beta}$ belong to the $u\left(n\right)$ subalgebra
of $so\left(2n\right)$. The $n$ operators $c_{\alpha\beta}$ for
$\alpha=\beta$ form the Cartan subalgebra of $so\left(2n\right)$.

The representation (\ref{eq:Hilbert-space-I}) is a $2^{n}$-dimensional
spinor representation of $so\left(2n\right)$, and its irreducible
representations are given by states with an even or odd number of
creation operators, corresponding to the $2^{n-1}$-dimensional Weyl
(semi-spinor) representations of $so\left(2n\right)$. These states
can be further decomposed in irreducible representations of $su\left(N\right)$,
since the isospin generators $\hat{t}_{a}$ are a linear combination
of the $so\left(2n\right)$ generators,\[
\hat{t}_{a}=t_{a\alpha\beta}a_{\alpha}^{\dagger}a_{\beta}=t_{a\alpha\beta}\left(-2c_{\beta\alpha}+\delta_{\alpha\beta}\right)=-2t_{a\alpha\beta}c_{\beta\alpha}\,.\]
 Therefore, we see that the $\hat{t}_{a}$ generate a $su\left(N\right)$
subalgebra of $so\left(2n\right)$.

In general, in order to determine the $SU\left(N\right)$ content
of the wave function, one proceeds as in \cite{Balachandran:1976ya}:
given $t_{a}$ an irreducible representation of $su\left(N\right)$
in terms of $n\times n$ matrices, the wave function belongs to a
(Weyl) semi-spinor representation of $so\left(2n\right)$. Then, one
decomposes the set of Cartan generators of $su\left(N\right)$ (a
maximal set of commuting generators) in terms of the $n$ Cartan generators
of $so\left(2n\right)$. In the special case of the representation
(\ref{eq:Hilbert-space-I}), one can choose the operators $c_{\alpha}=\left[a_{\alpha},a_{\alpha}^{\dagger}\right]/2$,
$\alpha=1,...,n$ as the maximum set of commuting generators of $so\left(2n\right)$.
For instance, in the case of $SU\left(2\right)$ one can take the
isospin projection $I_{1}$ and for $SU\left(3\right)$ one can take
the isospin projection $I_{1}$ and the hypercharge $Y$ to characterize
irreducible representations. One then decomposes isospin generators
in terms of the $c_{\alpha}$ to obtain their eigenvalues for the
spinor representation of $so\left(2n\right)$to which the wave function
belongs. The range of these eigenvalues gives the irreducible representations
of $SU\left(N\right)$. Therefore, to each given $n$-dimensional
irreducible representation $t_{a}$ of $SU\left(N\right)$, the wave
function will belong to a $2^{n-1}$-dimensional representation of
$SO\left(2n\right)$ (a semi-spinor representation), which decomposes
into irreducible representations of $SU\left(N\right)$ as determined
by the isospin generators $\hat{t}_{a}$.

In the special case of $SU\left(2\right)$, since it is of rank $1$,
the Cartan subalgebra is generated by a single element, say $t_{3}$,
whose matrix representation in a basis of isospin $s$ eigenstates
is of the form $t_{3}=\mathrm{diag}(s,s-1,...,-s+1,-s)$. The decomposition
in Cartan generators of $so\left(4s+2\right)$ of the isospin operator
$\hat{t}_{3}$ is as follows, \[
\hat{t}_{3}=sc_{1}+\left(s-1\right)c_{2}+\cdots+\left(-s\right)c_{2s+1}\,.\]
 Each $c_{\alpha}$ can take either of the values plus or minus $1/2$.
However, the wave function is in a state of either an even number
of plus $+1/2$ (even Weyl spinor) or an odd number of $+1/2$ (odd
Weyl spinor). For instance, for $s=1/2$, the possible eigenvalues
of $t_{3}$for the even representation is twice $0$, giving two scalar
representations; and for the odd representation is $\pm1/2$, giving
the isospin $1/2$ representation. For integer spin, even and odd
representations decompose in the same way, and the largest representation
is of spin $\left(s+1\right)s/2$. For example, $s=1$ gives the eigenvalues
$1,0,-1$ and again $0$, giving the isospin $1$ representation plus
a scalar. We summarize the results for some values of isospin in the
table below,

\begin{tabular}{|c|c|c|c|}
\hline 
isospin &
symmetry group &
representation dimension &
decomposition(even;odd) \tabularnewline
\hline
\hline 
$0$ &
$SO\left(2\right)$ &
$1$ &
$\underline{0}$ \tabularnewline
\hline 
$1/2$ &
$SO\left(4\right)$ &
$2$ &
$2\times\underline{0}$ ; $\underline{\frac{1}{2}}$ \tabularnewline
\hline 
$1$ &
$SO\left(6\right)$ &
$4$ &
$\underline{0}+\underline{1}$ \tabularnewline
\hline 
$3/2$ &
$SO\left(8\right)$ &
$8$ &
$3\times\underline{0}+\underline{2}$ ; $2\times\underline{\frac{3}{2}}$ \tabularnewline
\hline 
$2$ &
$SO\left(10\right)$ &
$16$ &
$\underline{0}+\underline{1}+\underline{2}+\underline{3}$ \tabularnewline
\hline
\end{tabular}

Thus, in order to obtain the fundamental representation of $SU\left(2\right)$
upon quantization, one must choose the Hilbert space to be the odd
Weyl spinor representation of $SO\left(4\right)$ of two-component
spinors. In this case, one gets from the constraint $T$ the Dirac
quantization condition \begin{equation}
\hat{T}\phi=\left[\left(\hat{p}_{\mu}+qA_{\mu}^{a}t_{a}\right)^{2}-m^{2}\right]\phi\left(x\right)=0\,,\label{eq:particular-coherentKG}\end{equation}
 which is precisely the wave equation in (\ref{flat-propagator-def})
for $t_{a}=\frac{1}{2}\sigma_{a}$.

It also possible to arrive at these results starting from the classical
action (\ref{eq:coherent-action}). In the following, it will be convenient
to express the Grassmann variables $\chi$ in terms of their real
and imaginary parts,\[
\chi_{\alpha}=\frac{1}{\sqrt{2}}\left(\chi_{1\alpha}+i\chi_{2\alpha}\right)\,,\]
 so that we are left with the real variables\begin{equation}
\chi_{1\alpha}=\frac{1}{\sqrt{2}}\left(\chi_{\alpha}+\bar{\chi}_{\alpha}\right)\,,\,\,\chi_{2\alpha}=\frac{1}{i\sqrt{2}}\left(\chi_{\alpha}-\bar{\chi}_{\alpha}\right)\,.\label{eq:real-algebra}\end{equation}
 In this way, the Grassmanian kinetic term becomes\[
L_{kin}=\frac{i}{4}\left(\chi_{1}\dot{\chi}_{1}-\dot{\chi}_{1}\chi_{1}+\chi_{2}\dot{\chi}_{2}-\dot{\chi}_{2}\chi_{2}\right)\,.\]
 $L_{kin}$ is invariant under transformations induced by $R_{\alpha\beta}=-i\left(\chi_{1\alpha}\chi_{1\beta}+\chi_{2\alpha}\chi_{2\beta}\right)$
and $S_{\alpha\beta}=-i\left(\chi_{1\alpha}\chi_{2\beta}+\chi_{1\beta}\chi_{2\alpha}\right)$
\begin{align*}
 & \delta_{\omega}\chi_{i\alpha}\equiv\left\{ \frac{1}{2}\omega_{\beta\gamma}R_{\beta\gamma},\chi_{i\alpha}\right\} _{D\left(\phi\right)}=\omega_{\alpha\beta}\chi_{i\beta}\,,\\
 & \delta_{\lambda}\chi_{i\alpha}\equiv\left\{ \frac{1}{2}\lambda_{\beta\gamma}S_{\beta\gamma},\chi_{i\alpha}\right\} _{D\left(\phi\right)}=\left(-1\right)^{i+1}\lambda_{\alpha\beta}\chi_{i\beta}\end{align*}
 where the Dirac brackets for the real variables follows from the
old variables' brackets (\ref{eq:canonical-brackets-I}) and their
expression in terms of the real variables (\ref{eq:real-algebra}),\[
\left\{ \chi_{1\alpha},\chi_{1\beta}\right\} _{D\left(\phi\right)}=\left\{ \chi_{2\alpha},\chi_{2\beta}\right\} _{D\left(\phi\right)}=-i\delta_{\alpha\beta}\,,\,\,\left\{ \chi_{1\alpha},\chi_{2\beta}\right\} _{D\left(\phi\right)}=0\,.\]
 The symmetry generators $R_{\alpha\beta}$ and $S_{\alpha\beta}$
satisfy the Lie algebra%
{} \begin{align*}
\left\{ R_{\alpha\beta},R_{\gamma\delta}\right\} _{D\left(\phi\right)} & =\delta_{\alpha\gamma}R_{\beta\delta}+\delta_{\beta\delta}R_{\alpha\gamma}-\delta_{\alpha\delta}R_{\beta\gamma}-\delta_{\beta\gamma}R_{\alpha\delta}\,,\\
\left\{ S_{\alpha\beta},S_{\gamma\delta}\right\} _{D\left(\phi\right)} & =\delta_{\alpha\gamma}R_{\beta\delta}+\delta_{\beta\delta}R_{\alpha\gamma}+\delta_{\alpha\delta}R_{\beta\gamma}+\delta_{\beta\gamma}R_{\alpha\delta}\,,\\
\left\{ R_{\alpha\beta},S_{\gamma\delta}\right\} _{D\left(\phi\right)} & =\delta_{\alpha\gamma}S_{\beta\delta}-\delta_{\beta\delta}S_{\alpha\gamma}+\delta_{\alpha\delta}S_{\beta\gamma}-\delta_{\beta\gamma}S_{\alpha\delta}\,.\end{align*}
 Above we recognize the commutation relations of the combination of
the $o\left(2n\right)$ generators $L_{ij}$, $i,j=1,...,2n$,\[
R_{\alpha\beta}=L_{2\alpha-1,2\beta-1}+L_{2\alpha,2\beta}\,,\,\, S_{\alpha\beta}=L_{2\alpha,2\beta-1}-L_{2\alpha-1,2\beta}\,-\delta_{\alpha\beta}\,.\]
 Moreover, from the following decomposition of the generators $I_{a}$
in terms of the symmetric and antisymmetric part of $t_{a}$,\begin{align}
I_{a} & =t_{a\left(\alpha\beta\right)}\left(\bar{\chi}_{\alpha}\chi_{\beta}+\bar{\chi}_{\beta}\chi_{\alpha}\right)+t_{a\left[\alpha\beta\right]}\left(\bar{\chi}_{\alpha}\chi_{\beta}-\bar{\chi}_{\beta}\chi_{\alpha}\right)\nonumber \\
 & =\frac{i}{2}t_{a\left(\alpha\beta\right)}\left(\chi_{1\alpha}\chi_{2\beta}+\chi_{1\beta}\chi_{2\alpha}\right)+\frac{1}{2}t_{a\left[\alpha\beta\right]}\left(\chi_{1\alpha}\chi_{1\beta}+\chi_{2\alpha}\chi_{2\beta}\right)\,,\nonumber \\
 & =-\frac{1}{2}t_{a\left(\alpha\beta\right)}S_{\alpha\beta}+\frac{i}{2}t_{a\left[\alpha\beta\right]}R_{\alpha\beta}\label{eq:isospin-decomp-I}\end{align}
 we again find the $I_{a}$ are a linear combination of $R_{\alpha\beta}$
and $S_{\alpha\beta}$, which is to say that the $I_{a}$ are the
generators of the subalgebra $su\left(N\right)$ of $so\left(2n\right)$.
It is $so\left(2n\right)$ and not $o\left(2n\right)$, because the
trace part of $S_{\alpha\beta}$ in the expansion of $I_{a}$ gives
no contribution, since the $t_{a}$ are traceless.

\section{Path integral in Clifford algebra representation}

\subsection{Path integral}

We use the representation (\ref{eq:antisym-gamma-rep}) for the generators
$T_{a}$ and standard techniques \cite{Fradkin:1991ci,Vasiliev:1998cq}
from the spinning particle case, adapted to our present problem, to
represent the causal propagator. In this case, the indices $\alpha$,
$\beta$ and $\gamma$ label the matrix entries of the $\Gamma$-matrices,
that is, they label the representation space for the Clifford algebra.
The propertime representation for the operator $\hat{D}$ (\ref{eq:propertime})
in the position representation is \begin{equation}
D\left(x_{out},x_{in}\right)=i\int_{0}^{\infty}\left\langle x_{out}\right|e^{-i\hat{H}\left(\lambda\right)}\left|x_{in}\right\rangle d\lambda\,.\label{proper-time-representation-matrix-element}\end{equation}
 Next a discretization is made inserting $N-1$ identity resolutions
$I=\int dx\left|x\right\rangle \left\langle x\right|$ in the above
expression,\begin{align}
D\left(x_{out},x_{in}\right) & =\lim_{N\rightarrow\infty}i\int_{0}^{\infty}d\lambda_{0}\int_{-\infty}^{\infty}\left(\prod_{i=1}^{N-1}dx_{i}\right)d\lambda_{1}\cdots d\lambda_{N}\nonumber \\
 & \prod_{i=1}^{N}\left\langle x_{i}\right|e^{-i\hat{H}\left(\lambda_{i}\right)/N}\left|x_{i-1}\right\rangle \delta\left(\lambda_{i}-\lambda_{i-1}\right)\label{eq:propagator-discretization2}\end{align}
 where $x_{N}=x_{out}$ and $x_{0}=x_{in}$. Applying the symmetric
or Weyl correspondence to the general matrix element, one has\begin{align}
\left\langle x_{i}\right|e^{-i\hat{H}\left(\lambda_{i}\right)/N}\left|x_{i-1}\right\rangle  & =\int\frac{dp_{i}}{\left(2\pi\right)^{4}}\exp\left(-\frac{i}{N}H\left(\lambda_{i},\frac{x_{1}+x_{2}}{2},p_{i}\right)\right)e^{i\left(x_{i}-x_{i-1}\right)p_{i}}\,,\label{general-matrix-element2}\end{align}
 where $H$ is the Weyl symbol of $\hat{H}$,\[
H\left(\lambda,x,p\right)=\lambda\left[m^{2}-\left(p_{i}^{2}+qp_{i}^{\mu}A_{\mu}^{a}\left(x\right)T_{a}\right)^{2}\right]\,.\]
 As in the spinning particle case \cite{Fradkin:1991ci}, one assigns
to each matrix $T_{a}$ its 'time' $\tau_{j}=j\Delta\tau$, so that
the time-ordered (\ref{eq:propagator-discretization2}) becomes, for
$1/N\equiv\Delta\tau$,\begin{align}
D\left(x_{out},x_{in}\right) & =\lim_{\Delta\tau\rightarrow0}iT\int_{0}^{\infty}d\lambda\int_{-\infty}^{\infty}\left(\prod_{i=1}^{N-1}dx_{i}\right)\left(\prod_{i=1}^{N}\frac{\mathbf{\mathrm{d}}p_{i}}{\left(2\pi\right)^{4}}d\lambda_{i}\frac{d\pi_{i}}{2\pi}\right)\nonumber \\
 & \times\exp i\sum_{i=1}^{N}S_{i}\left(x_{i},x_{i-1},p_{i},\lambda_{i},\pi_{i}\right)\,,\label{discrete-path-integral}\end{align}
 where\begin{equation}
S_{i}=\left(\frac{x_{i}-x_{i-1}}{\Delta\tau}\cdot p_{i}-H\left(\lambda_{i},\frac{x_{i}+x_{i-1}}{2},p_{i}\right)+\pi_{i}\frac{\lambda_{i}-\lambda_{i-1}}{\Delta\tau}\right)\Delta\tau\,.\label{discrete-hamiltonian-action}\end{equation}
 In the limit $\Delta\tau\rightarrow0$, $S_{i}\rightarrow S_{H}\left[x,p;\tau_{in},\tau_{out}\right]$
is the Hamiltonian action, a functional of the trajectory $\left(x\left(t\right),p\left(t\right)\right)$
in phase space, defined in the proper-time interval $\left[\tau_{in},\tau_{out}\right]$,
and (\ref{discrete-path-integral}) is the discrete version of the
following path integral in the Hamiltonian form:\begin{equation}
D\left(x_{out},x_{in}\right)=iT\int_{0}^{\infty}d\lambda_{0}\int_{x_{in}}^{x_{out}}Dx\int Dp\int_{\lambda_{0}}D\lambda D\pi\exp i\int_{\tau_{in}}^{\tau_{out}}\left(\dot{x}\cdot p-H\left(\lambda,x,p\right)+\pi\dot{\lambda}\right)d\tau\,.\label{hamiltonian-path-integral}\end{equation}
 Following \cite{Fradkin:1991ci}, we introduce odd sources $\rho_{a}\left(\tau\right)$,
anticommuting with the $\Gamma$-matrices, and rewrite \ref{hamiltonian-path-integral}
as\begin{align*}
D\left(x_{out},x_{in}\right) & =i\int_{0}^{\infty}d\lambda_{0}\int_{x_{in}}^{x_{out}}Dx\int Dp\int_{\lambda_{0}}D\lambda D\pi\exp i\int_{0}^{1}\left[\lambda\left(\left(p_{\mu}+\frac{q}{4}t_{a\beta}^{\alpha}A_{\mu}^{a}\frac{\delta_{l}}{\delta\rho_{\alpha}}\frac{\delta_{l}}{\delta\rho_{\beta}}\right)^{2}-m^{2}\right)\right.\\
 & \left.\left.p\cdot\dot{x}+\pi\dot{\lambda}d\tau\right]\times T\int_{0}^{1}\rho_{\alpha}\left(\tau\right)\Gamma^{\alpha}d\tau\right|_{\rho=0}\,,\end{align*}
 where for simplicity we have made $\tau_{in}=0$ and $\tau_{out}=1$.
It is possible to present the last term on the right-hand side of
the above equation as a path integral \cite{Fradkin:1991ci,Vasiliev:1998cq},\begin{align*}
 & T\int_{0}^{1}\rho_{\alpha}\left(\tau\right)\Gamma^{\alpha}d\tau=\exp\left(i\Gamma^{\alpha}\frac{\partial_{l}}{\partial\theta^{\alpha}}\right)\\
 & \times\left.\underset{\psi\left(0\right)+\psi\left(1\right)=\theta}{\int}\exp\left[\int_{0}^{1}\left(\psi^{\alpha}\left(\tau\right)\dot{\psi}_{\alpha}\left(\tau\right)-i2\rho_{\alpha}\left(\tau\right)\psi^{\alpha}\left(\tau\right)\right)d\tau+\psi^{\alpha}\left(1\right)\psi_{\alpha}\left(0\right)\right]\mathcal{D}\psi\right|_{\theta=0}\\
 & \mathcal{D}\psi=D\psi\left[\underset{\psi\left(0\right)+\psi\left(1\right)=0}{\int}\exp\int_{0}^{1}\psi^{\alpha}\left(\tau\right)\dot{\psi}_{\alpha}\left(\tau\right)d\tau\right]^{-1}\,,\end{align*}
 where $\theta$ are odd constants, anticommuting with the $\Gamma$-matrices.
Then, we arrive at the Hamiltonian path-integral representation for
the propagator:\begin{align*}
D\left(x_{out},x_{in}\right) & =i\exp\left(i\Gamma^{\alpha}\frac{\partial_{l}}{\partial\theta^{\alpha}}\right)\int_{0}^{\infty}d\lambda_{0}\int_{x_{in}}^{x_{out}}Dx\int Dp\int_{\lambda_{0}}D\lambda D\pi\\
 & \int\exp\left\{ i\int_{0}^{1}\left[\lambda\left(\left(p_{\mu}-qt_{a\beta}^{\alpha}A_{\mu}^{a}\psi_{\alpha}\psi_{\beta}\right)^{2}-m^{2}\right)\right.\right.\\
 & \left.\left.\left.-i\psi^{\alpha}\dot{\psi}_{\alpha}+p\cdot\dot{x}+\pi\dot{\lambda}\right]d\tau+\psi^{\alpha}\left(1\right)\psi_{\alpha}\left(0\right)\right\} \mathcal{D}\psi\right|_{\theta=0}\,,\\
 & x\left(0\right)=x_{in}\,,\,\, x\left(1\right)=x_{out}\,,\,\,\lambda\left(0\right)=\lambda_{0}\,,\,\,\psi\left(0\right)+\psi\left(1\right)=\theta\,.\end{align*}
 Integrating over the momenta, one finds the Lagrangian path-integral
representation:\begin{align}
D\left(x_{out},x_{in}\right) & =i\exp\left(i\Gamma^{\alpha}\frac{\partial_{l}}{\partial\theta^{\alpha}}\right)\int_{0}^{\infty}de_{0}\int\exp\left\{ i\left(S_{eff}+S_{G}\right)+\psi^{\alpha}\left(1\right)\psi_{\alpha}\left(0\right)\right\} \left.M\left[e,x\right]DxDeD\pi\mathcal{D}\psi\right|_{\theta=0}\nonumber \\
 & S_{eff}=i\int_{0}^{1}\left(-\frac{\dot{x}^{2}}{2e}-\frac{e}{2}m^{2}+qt_{a\beta}^{\alpha}\dot{x}^{\mu}A_{\mu}^{a}\psi_{\alpha}\psi_{\beta}-i\psi^{\alpha}\dot{\psi}_{\alpha}\right)\nonumber \\
 & x\left(0\right)=x_{in}\,,\,\, x\left(1\right)=x_{out}\,,\,\, e\left(0\right)=e_{0}\,,\,\,\psi\left(0\right)+\psi\left(1\right)=\theta\,,\label{eq:lagrangian-pathintegral2}\end{align}
 where the measure $M\left[e,x\right]$ and $S_{G}$ are \[
M\left[e,x\right]=\int Dp\exp\frac{i}{2}\int_{0}^{1}ep^{2}d\tau\,,\,\, S_{G}=\int_{0}^{1}\pi\dot{e}\,.\]

\subsection{Pseudoclassical action}

Let us consider the reparametrization invariant action from (\ref{eq:lagrangian-pathintegral2})
with the rescaling $\psi\rightarrow i/\sqrt{2}\psi$,\begin{equation}
S_{eff}=\int dx^{4}\left(-\frac{\dot{x}^{2}}{2e}-\frac{e}{2}m^{2}-q\dot{x}^{\mu}A_{\mu}^{a}I_{a}+\frac{i}{2}\psi^{\alpha}\dot{\psi}_{\alpha}\right)\,,\,\, I_{a}=\frac{1}{2}t_{\alpha\beta}^{\alpha}\psi_{\alpha}\psi_{\beta}\,.\label{eq:clifford-action}\end{equation}
 The above action is essentially the one written in \cite{Barducci:1976xq}
in the case the set of Grassmann variables $\psi$ belong to the adjoint
representation of a compact simple group $G$ ($t_{a\beta}^{\alpha}=f_{ac}^{b}$),
and in \cite{Balachandran:1976ya} for $\psi$ in a representation
with antisymmetric generators $t_{a}$. The equations of motion in
the gauge $e=\sqrt{\dot{x}^{2}}/m$ are\begin{align*}
 & \frac{d}{dt}\left(m\frac{\dot{x}^{\mu}}{\sqrt{\dot{x}^{2}}}\right)=qF_{\mu\nu}^{a}\dot{x}^{\nu}I_{a}\,,\,\, D_{t}\psi^{\alpha}\equiv\frac{d}{dt}\psi^{\alpha}+iq\dot{x}^{\mu}A_{\mu}^{a}t_{a\beta}^{\alpha}\psi^{\beta}=0\,,\\
 & F_{\mu\nu}^{a}=\partial_{\mu}A_{\nu}^{a}-\partial_{\nu}A_{\mu}^{a}+iqf_{bc}^{a}A_{\mu}^{b}A_{\nu}^{c}\,,\end{align*}
 where $F_{\mu\nu}^{a}$ is the non-Abelian field strength.

Next, we follow a similar canonical analysis path than the one taken
in the case of the coherent representation, this time with $t_{a}$
denoting $n\times n$ antisymmetric matrices. As expected, the Hamiltonian
is proportional to constraints, \[
H=-\frac{e}{2}T-\dot{\psi}^{\alpha}\phi_{\alpha}\]
 where \[
T=\left(p_{\mu}+qA_{\mu}^{a}I_{a}\right)^{2}-m^{2}\,,\,\,\phi_{\alpha}=\pi_{\alpha}-\frac{i}{2}\psi_{\alpha}\]
 After redefining $T$ through the shift $\psi\rightarrow\psi-i\phi$,
$T\rightarrow\tilde{T}$, the constraint algebra becomes\[
\left\{ \tilde{T},\phi_{\alpha}\right\} =0\,,\,\,\left\{ \phi_{\alpha},\phi_{\beta}\right\} =-i\delta_{\alpha\beta}\,.\]
 The set $\Phi=\left\{ \tilde{T},\phi\right\} $ is first-class, and
the evolution of the independent variables $\eta=\left(x,p,\psi\right)$
is\[
\dot{\eta}=\left\{ \eta,\Lambda T\right\} _{D\left(\phi\right)}=0\,,\,\, T=\phi_{\alpha}=0\,,\]
 where the Dirac brackets are defined with respect to the second-class
constraint set $\left\{ \phi\right\} $. The Dirac commutator of the
independent variables is\[
\left\{ x^{\mu},p_{\nu}\right\} _{D\left(\phi\right)}=\delta_{\nu}^{\mu}\,,\,\,\left\{ \psi_{\alpha},\psi_{\beta}\right\} _{D\left(\phi\right)}=-i\delta_{\alpha\beta}\,,\]
 The isospin quantities $I_{a}$ satisfy the Lie algebra of $SU\left(N\right)$
after quantization and are covariantly constant:\[
\left\{ I_{a},I_{b}\right\} _{D\left(\phi\right)}=-if_{ab}^{c}I_{c}\,,\,\, D_{\tau}I^{c}=\frac{d}{d\tau}I^{c}+iq\dot{x}^{\mu}A_{\mu}^{a}f_{ab}^{c}I^{b}=0\,.\]

It is clear that upon quantization the Grassmann variables $\psi_{\alpha}$
generate Clifford algebra with $n$ generators and positive-definite
inner product. And thus the physical states $\phi$ are $2^{[n/2]}$-component
vectors satisfying\begin{equation}
\left[\left(\hat{p}_{\mu}+qA_{\mu}^{a}\hat{I}_{a}\right)^{2}-m^{2}\right]\phi\left(x\right)=0\,,\label{eq:cliffordKG}\end{equation}
 where the quantum-mechanical isospin operators $\hat{I}_{a}=\frac{1}{4}t_{a\beta}^{\alpha}\Gamma_{\alpha}\Gamma_{\beta}$
are precisely those introduced in (\ref{eq:antisym-gamma-rep}) and
they satisfy the $su\left(N\right)$ algebra (\ref{eq:lie-algebra})\[
\left[\hat{I}_{a},\hat{I}_{b}\right]=f_{ab}^{c}\hat{I}_{c}\,.\]

Let us draw a similar analysis of the isospin content for the classical
theory as the one given in section 3.1. Here, the Grassmannian kinetic
terms in the action are invariant under the transformations generated
by $R_{\alpha\beta}=-i\psi_{\alpha}\psi_{\beta}$,\[
\delta_{\omega}\psi_{\alpha}=\left\{ \frac{1}{2}\omega_{\beta\gamma}R_{\beta\gamma},\psi_{\alpha}\right\} _{D\left(\phi\right)}=\omega_{\alpha\beta}\psi_{\beta}\,,\,\,\omega_{\alpha\beta}=-\omega_{\beta\alpha}\,,\]
 which give a representation for the Lie algebra $so\left(n\right)$:\[
\left\{ R_{\alpha\beta},R_{\gamma\delta}\right\} _{D\left(\phi\right)}=\delta_{\alpha\gamma}R_{\beta\delta}-\delta_{\beta\gamma}R_{\alpha\delta}-\delta_{\alpha\delta}R_{\beta\gamma}+\delta_{\beta\delta}R_{\alpha\gamma}\,.\]
 So the the generators $I_{a}$ are a linear combination of the generators
$L_{\alpha\beta}$ of $so\left(n\right)$, and therefore they generate
an $su\left(N\right)$ subalgebra of $so\left(n\right)$. In order
to determine the $SU\left(N\right)$ content of the wave function,
one proceeds as in \cite{Balachandran:1976ya}: given $t_{a}$ an
irreducible representation of $su\left(N\right)$ in terms of $n\times n$
\textbf{antisymmetric} matrices, it is clear \textbf{$\psi_{\alpha}$}
gives rise to a Clifford algebra in the quantum theory, so the wave
function can be taken to belong to a spinor representation of $so\left(n\right)$.
One then calculates the eigenvalues of a maximal set of commuting
generators $I_{a}$ for this representation, and thus determines how
it decomposes in irreducible representations of $SU\left(N\right)$.
Therefore, to each given $n$-dimensional irreducible representation
$t_{a}$ of $SU\left(N\right)$, the wave function will belong to
a $2^{[n/2]}$-dimensional representation of $SO\left(n\right)$,
which decomposes into irreducible representations of $SU\left(N\right)$
as determined by the isospin generators $I_{a}$.

For example, in the case of $SU\left(2\right)$, one can find a basis
for which $t_{a}$ are antisymmetric, and in which $I_{1}$ decomposes
as\[
I_{1}=L_{23}+2L_{45}+\cdots+sL_{2s,2s+1}\,.\]
 The wave function is a $2^{s}$-component spinor of $SO\left(2s+1\right)$.
Below we give the $SU\left(2\right)$ decomposition of the spinor
$SO\left(2s+1\right)$ representation for some values of isospin:

\begin{tabular}{|c|c|c|c|}
\hline 
isospin &
symmetry group &
representation dimension &
decomposition \tabularnewline
\hline
\hline 
$1$ &
$SO\left(3\right)$ &
$2$ &
$\underline{\frac{1}{2}}$ \tabularnewline
\hline 
$2$ &
$SO\left(5\right)$ &
$4$ &
$\underline{\frac{3}{2}}$ \tabularnewline
\hline 
$3$ &
$SO\left(7\right)$ &
$8$ &
$\underline{0}+\underline{3}$ \tabularnewline
\hline
\end{tabular}

\section{Summary}

We have described two methods of generating classical actions for
a scalar particle with isospin via path-integral representations of
the causal propagator. Dirac quantization of these actions produce
the corresponding wave equations for various possible representations
of $SU\left(N\right)$. By means of a judicious choice of the pseudo-classical
action and the representation of the $su\left(N\right)$ algebra in
the action, it is possible to obtain the wave action for any desired
isospin.

\begin{description}
\item [{Acknowledgment:}] R.F. thanks FAPESP for support and D.M.G acknowledges
FAPESP and CNPq for permanent support. 
\end{description}
\appendix

\section{Weyl ordering of operators and functions in the Berezin algebra}

Let us write the Hamiltonian operator (\ref{eq:propertime}) explicitly:\[
\hat{H}=-\lambda\left(\hat{p}^{2}+qt_{a\alpha\beta}\left(\hat{p}^{\mu}\hat{A}_{\mu}^{a}+\hat{A}_{\mu}^{a}\hat{p}^{\mu}\right)\hat{a}_{\alpha}^{\dagger}\hat{a}_{\beta}+q^{2}t_{a\alpha\beta}t_{b\gamma\delta}\hat{A}_{\mu}^{a}\hat{A}^{\mu b}\hat{a}_{\alpha}^{\dagger}\hat{a}_{\beta}\hat{a}_{\gamma}^{\dagger}\hat{a}_{\delta}-m^{2}\right)\,.\]
 Total symmetrization in $\hat{x}$and $\hat{p}$, and total antisymmetrization
in $a^{\dagger}$ and $a$ gives the Weyl-ordered Hamiltonian operator
$\hat{H}_{W}$:\begin{equation}
\hat{H}_{W}=-\lambda\left(\hat{p}^{2}+\frac{q}{2}t_{a\alpha\beta}\left(\hat{p}^{\mu}\hat{A}_{\mu}^{a}+\hat{A}_{\mu}^{a}\hat{p}^{\mu}\right)\left[\hat{a}_{\alpha}^{\dagger},\hat{a}_{\beta}\right]+q^{2}t_{a\alpha\beta}t_{b\gamma\delta}\hat{A}_{\mu}^{a}\hat{A}^{\mu b}\left(\hat{a}_{\alpha}^{\dagger}\hat{a}_{\beta}\hat{a}_{\gamma}^{\dagger}\hat{a}_{\delta}\right)_{W}-m^{2}\right)\,,\label{eq:hamiltonian-wo}\end{equation}
 where the four-fermion term is given by\[
\hat{a}_{\alpha}^{\dagger}\hat{a}_{\beta}\hat{a}_{\gamma}^{\dagger}\hat{a}_{\delta}=\left(\hat{a}_{\alpha}^{\dagger}\hat{a}_{\beta}\hat{a}_{\gamma}^{\dagger}\hat{a}_{\delta}\right)_{W}+\frac{1}{2}\delta_{\gamma\delta}\left(\hat{a}_{\alpha}^{\dagger}\hat{a}_{\beta}\right)_{W}-\frac{1}{2}\delta_{\delta\alpha}\left(\hat{a}_{\gamma}^{\dagger}\hat{a}_{\beta}\right)_{W}+\delta_{\alpha\beta}\left(\hat{a}_{\gamma}^{\dagger}\hat{a}_{\delta}\right)_{W}+\frac{1}{2}\delta_{\gamma\beta}\left(\hat{a}_{\alpha}^{\dagger}\hat{a}_{\delta}\right)_{W}-\frac{1}{4}\delta_{\delta\alpha}\delta_{\gamma\beta}-\delta_{\alpha\beta}\delta_{\gamma\delta}\]
 Using the tracelessness of the matrices $t_{a}$ and antisymmetry
of the structure constants $f_{abc}$, we have\[
\hat{H}=\hat{H}_{W}+\lambda\frac{q^{2}}{4}\mathrm{tr}\left(t_{a}t_{b}\right)\hat{A}_{\mu}^{a}\hat{A}^{\mu b}\]
 Thus, the Hamiltonian is the sum of a Weyl-ordered expression plus
a gauge non-invariant contribution. The Weyl-symbol corresponding
to $\hat{H}_{W}$ is \begin{equation}
H_{W}=-\lambda\left(p^{2}+2qt_{a\alpha\beta}\left(p^{\mu}A_{\mu}^{a}\right)\bar{\chi}_{\alpha}\chi_{\beta}+q^{2}t_{a\alpha\beta}t_{b\gamma\delta}A_{\mu}^{a}A^{\mu b}\bar{\chi}_{\alpha}\chi_{\beta}\bar{\chi}_{\gamma}\chi_{\delta}-m^{2}\right)\label{eq:weyl-symbol}\end{equation}

\subsection[]{Proof%
\footnote{Adapted from \cite{gavazzi89}%
} of fermionic midpoint rule.}

If $F\left(\hat{a},\hat{a}^{\dagger}\right)$ is any Weyl-ordered polynomial in
$\hat{a}$ and $\hat{a}^{\dagger}$, then\begin{align}
\left\langle \bar{\chi}\right|F\left(\hat{a},\hat{a}^{\dagger}\right)\left|\chi\right\rangle  & =\int d\bar{\eta}d\eta\left\langle \bar{\chi}\right|\left.\eta\right\rangle F\left(\frac{\chi+\eta}{2},\bar{\eta}\right)\left\langle \bar{\eta}\right.\left|\chi\right\rangle \,,\label{eq:fermionic-midpoint}\\
 & =\int d\bar{\eta}d\eta\left\langle \bar{\chi}\right|\left.\eta\right\rangle F\left(\chi,\frac{\bar{\chi}+\bar{\eta}}{2}\right)\left\langle \bar{\eta}\right.\left|\chi\right\rangle \,,\label{eq:fermionic-midpoint2}\end{align}
 Let us prove the identity (\ref{eq:fermionic-midpoint}). The proof
of the second identity is analogous. First, consider $F\left(\hat{a}^{\dagger}\right)$
a polynomial in creation operators. Clearly, $F$ is Weyl-ordered,
and (\ref{eq:fermionic-midpoint}) is trivially satisfied, \[
\left\langle \bar{\chi}\right|F\left(\hat{a}^{\dagger}\right)\left|\chi\right\rangle =\int d\bar{\eta}d\eta\left\langle \bar{\chi}\right|\left.\eta\right\rangle F\left(\bar{\chi}\right)\left\langle \bar{\eta}\right.\left|\chi\right\rangle \,.\]
 Now, for $F\left(\hat{a},\hat{a}^{\dagger}\right)=\frac{1}{2}\left(\hat{a}_{\alpha}f\left(\hat{a}^{\dagger}\right)+\left(-1\right)^{\varepsilon\left(f\right)}f\left(\hat{a}^{\dagger}\right)\hat{a}_{\alpha}\right)$,
where the plus or minus sign depends on the parity of $f\left(\hat{a}^{\dagger}\right)$,
(\ref{eq:fermionic-midpoint}) is easily seen to hold. Any Weyl-ordered
polynomial can be obtained by repeated antisymmetrizations of the
form $F=\frac{1}{2}\left(\hat{a}_{\alpha}f\pm f\hat{a}_{\alpha}\right)$
where $f\left(\hat{a},\hat{a}^{\dagger}\right)$ is Weyl-ordered.
Therefore, let us prove (\ref{eq:fermionic-midpoint}) inductively,
by assuming it holds for $f\left(\hat{a},\hat{a}^{\dagger}\right)$
and proving it is also true for $F=\frac{1}{2}\left(\hat{a}_{\alpha}f\pm f\hat{a}_{\alpha}\right)$,
\begin{align*}
\left\langle \bar{\chi}\right|\frac{1}{2}\left(\hat{a}_{\alpha}f\pm f\hat{a}_{\alpha}\right)\left|\chi\right\rangle  & =\int d\bar{\eta}d\eta\frac{1}{2}\left(\left\langle \bar{\chi}\right|\hat{a}_{\alpha}\left|\eta\right\rangle \left\langle \bar{\eta}\right|f\left|\chi\right\rangle \pm\left\langle \bar{\chi}\right|\left.\eta\right\rangle \left\langle \bar{\eta}\right|f\hat{a}_{\alpha}\left|\chi\right\rangle \right)\\
 & =\int d\bar{\eta}d\eta\left\langle \bar{\chi}\right|\left.\eta\right\rangle \frac{\eta_{\alpha}+\chi_{\alpha}}{2}\left\langle \bar{\eta}\right|f\left|\chi\right\rangle \\
 & =\int d\bar{\eta}d\eta d\bar{\xi}d\xi\left\langle \bar{\chi}\right|\left.\eta\right\rangle \left\langle \bar{\eta}\right|\left.\xi\right\rangle \frac{\chi_{\alpha}+\eta_{\alpha}}{2}f\left(\frac{\chi+\xi}{2},\bar{\xi}\right)\left\langle \bar{\xi}\right.\left|\chi\right\rangle \\
 & =d\bar{\xi}d\xi\left\langle \bar{\chi}\right|\left.\xi\right\rangle \frac{\chi_{\alpha}+\xi_{\alpha}}{2}f\left(\frac{\chi+\xi}{2},\bar{\xi}\right)\left\langle \bar{\xi}\right.\left|\chi\right\rangle \end{align*}
 where in the last equality we used the identity \begin{equation}
\int d\bar{\eta}d\eta\left\langle \bar{\alpha}\right|\left.\eta\right\rangle \left\langle \bar{\eta}\right|\left.\beta\right\rangle f\left(\eta\right)=\left\langle \bar{\alpha}\right|\left.\beta\right\rangle f\left(\beta\right)\,.\label{eq:coherent-delta}\end{equation}

\bibliographystyle{unsrt}
\bibliography{bibtex}

\end{document}